\documentclass[aps,groupedaddress,showpacs,%
amsmath,preprintnumbers,onecolumn]{revtex4}

\usepackage{graphicx}
\usepackage{amsfonts}
\usepackage{amssymb}
\usepackage{amsbsy}
\usepackage{amsmath}
\usepackage{latexsym}
\usepackage{bm}
\usepackage{color}

\def\be {\begin{equation}}
\def\ee  {\end{equation}}
\def\bea {\begin{eqnarray}}
\def\eea {\end{eqnarray}}

\def\be {\begin{equation}}
\def\ee  {\end{equation}}
\def\bea {\begin{eqnarray}}
\def\eea {\end{eqnarray}}

\begin{document}
\preprint{ }
\title{Background independent quantization and the uncertainty principle}
\author{Golam Mortuza Hossain}
\email{ghossain@unb.ca}
\author{Viqar Husain}
\email{vhusain@unb.ca}
\author{Sanjeev S.~Seahra}
\email{sseahra@unb.ca } \affiliation{ Department of Mathematics and
Statistics, University of New Brunswick, Fredericton, NB, Canada E3B
5A3\\} \pacs{04.60.Ds}
\date{\today}
\begin{abstract}

It is shown that polymer quantization leads to a modified uncertainty
principle similar to that obtained from string theory and non-commutative
geometry. When applied to quantum field theory  on general background
spacetimes, corrections to the uncertainty principle acquire a  metric
dependence. For Friedmann-Robertson-Walker cosmology this translates to a
scale factor dependence which gives a large effect in the  early universe.

 \end{abstract}

\maketitle

\section{Introduction}

There is a belief that if  quantum gravity effects are
taken into account then the uncertainty principle of quantum mechanics is
modified. This is based on arguments  that make use of the existence of
a fundamental length scale in quantum gravity, and the possibility of black
hole formation at sufficiently large densities.  Candidate theories of quantum
gravity such as string theory and non-commutative spacetime models provide
a more concrete realization of this possibility. 

An early suggestion that gravity  has the possibility of affecting
the uncertainty principle  was given by Mead  \cite{mead}, who
re-considered the Heisenberg microscope experiment with
gravitational interaction. The basic argument does not require
quantum gravity, and can be made using Newtonian gravity along with
special relativity: Consider a probe photon of frequency
$\omega$ and a  particle of mass $m$ that is to be observed. The
acceleration  of the particle due to gravity when the probe is
within a distance $\Delta x$ is $a =G\omega/(\Delta x)^2$, where $G$
is Newton's constant. As a result its gravitationally induced
displacement in a time interval $\Delta t $ is
\be \label{GedExpResult}
\Delta x  \ge a (\Delta t)^2 \sim G\omega \sim G \Delta p,
\ee
where the last step follows because the particle's
momentum is uncertain by an amount given by the probe momentum. Thus
including gravity suggests an additional contribution to position
uncertainty that is proportional to the momentum uncertainty.  A more
refined general relativistic view of this gedanken experiment gives a
similar result. 
 
Beyond gedanken experiments, modified uncertainty relations can be
derived from candidate theories of quantum gravity that come with a
fundamental length scale. For example, in string theory a basic
observation is that the mass of a string is proportional to its
length. Strings therefore expand in size when probed at sufficiently
high energies. This suggests an additional uncertainty in 
the position of a string which is proportional to the uncertainty in its
momentum $\Delta p$. This indicates a modification to the uncertainty
relation of type \cite{Yoneya, Amati}
\be
 \Delta x \ge  \frac{1}{2\Delta p}  + k\  l_P^2 \Delta p,
 \label{mup}
\ee
where $l_P$ is the Planck length and $k$ is a dimensionless constant. The
second term in the expression (\ref{mup}) is of the same form as in the
equation (\ref{GedExpResult}). The generalized uncertainty relation
(\ref{mup}) can also be seen to arise from a theory independent
consideration of particle emission from a black hole \cite{maggiore-93};
the argument uses the observation that  the uncertainty in the horizon
radius is proportional to the emitted particle's mass.

A further observation follows from combining the relation $\Delta E \sim \Delta x$ for a string
with the usual time-energy uncertainty relation. This leads  to  an effective
non-commutativity of spacetime as first observed by Yoneya \cite{Yoneya,
Yoneya-rev}: 
\be
   \Delta x \Delta t \ge l_P^2. \label{xt-uncertainty}
\ee

The realization that string theory suggests a non-commutativity of
spacetime has been a motivation for postulating non-commutative
spacetime theories for the purpose of generalizing quantum field
theory and for developing models for quantum gravity. There are many
approaches for developing this idea, but  perhaps the main one
involves a dynamics independent  replacement of a classical manifold
by a non-commutative  one \cite{connes:2000, madore:book,
doplicher95}.  In this approach the commutative algebra of smooth
complex valued functions with specified fall-off conditions is
replaced by a non-commutative algebra, and much work has gone into
what algebra this should be. A related approach is to use modified
Heisenberg  \cite{Kempf-UP} and  $\kappa$-deformed Poincare algebras
\cite{MaggioreQG} as a basis for constructing quantum theory. This
approach has been applied to obtain corrections to atomic  spectra
\cite{Das-UP}.

In this work we consider an alternative approach based on polymer
quantization\cite{polymer}. In this mathematically well defined
quantization, the Hilbert space which is used for representing physical
variables as operators is significantly different from the usual
$L_2(\mathbb{R})$ (the space of square-integrable
functions). Unlike the approaches mentioned earlier, there is no deformation of
the algebra of observables; rather the Hilbert space is such that the
momentum operator is realized only indirectly as a function of  translation
operators. A direct consequence of this feature is that  the quantization 
comes with a fixed mass or length scale. This scale does not arise
from gravitational interaction as in the case of gedanken experiment.
However, its presence is crucial in modifying the uncertainty principle.

It has been noted in \cite{polymer} that the uncertainty principle receives
polymer corrections. Our formulation, however, provides a direct comparison
with the general form (\ref{mup}), and an extension to the field theory
where metric dependence of the correction terms becomes manifest.
This property may have interesting applications in the cosmological
settings where polymer corrections become significant in early universe. We
note further that when combined with the time-energy uncertainty relation,
polymer quantization provides a hint of space non-commutativity.

\section{Polymer quantization and the uncertainty relation}

Polymer quantization is a theory independent procedure that requires an
additional mass scale besides Planck's constant $\hbar$. As such it may be
viewed as a generalization of quantum mechanics where the
``Schr\"odinger'' or ``long wavelength'' limit is recovered  if  a suitably
defined dimensionless parameter is small. (For example, in polymer
quantization of harmonic oscillator of mass $m$ and frequency $\omega$ such
a parameter is $g= m\omega/M_\star^2$, where $M_\star$ denotes the polymer
mass scale.) More generally such a parameter could involve not just
coupling constants in a Hamiltonian but also parameters in a wave function
such as the width of a Gaussian wave packet.

The following summary of polymer quantization follows that given in
\cite{HLW-hyd}. We start with the basis states
\be
\psi_{x_0}(x) = \left \{ \begin{array}{r l} 1, & x=x_0 \\
0, & x\ne x_0 \ . \end{array}
\right .
\label{eq:basisstates}
\ee
The polymer Hilbert space is the Cauchy completion of the linear span
of these basis states in the inner product
\be
\langle \psi_x , \psi_{x'} \rangle = \delta_{x,x'},
\label{eq:ip}
\ee
where the quantity on the right-hand side is the Kronecker delta rather
than the Dirac delta. This Hilbert space is non-separable, and
inequivalent to $L_2(\mathbb{R})$  \cite{velhinho,corduneau,hewitt-ross}. Intuitively, constructing
a single nonzero $L_2(\mathbb{R})$ state using the polymer basis states,
would require an uncountable superposition, and thus leads to a
non-normalizable state in the polymer Hilbert space. Conversely, any state
in the polymer Hilbert space has support on at most countably many points,
and will thus represent the null state in $L_2(\mathbb{R})$.

Next we define the action of the basic quantum operators. The position
operator $\hat{x}$ acts by multiplication,
\be
\bigl(\hat{x} \psi \bigr) (x) =  x \psi(x),
\label{eq:xhat-action}
\ee
and its domain contains the linear span of the basis
states~(\ref{eq:basisstates}). The translation operators
$\hat{U}_{\lambda}$, $\lambda\in\mathbb{R}$, act by
\be
\bigl( \hat{U}_{\lambda} \psi \bigr) (x)
= \psi(x + \lambda) ,
\label{eq:Uhat-action}
\ee
and are clearly unitary. The parameter $\lambda$ is dimensionful
and defines the polymer mass scale. Formulas (\ref{eq:xhat-action}) and
(\ref{eq:Uhat-action}) are identical to those in $L_2(\mathbb{R})$.
However, in $L_2(\mathbb{R})$, the action of $\hat{U}_{\lambda}$ is weakly
continuous in~$\lambda$, and there exists a densely-defined self-adjoint
momentum operator $\hat{p}$ such that
$\hat{p} = -i
\bigl[\partial_\lambda
\hat{U}_{\lambda}\bigr]_{\lambda=0}
= i \partial_x$ and $\hat{U}_{\lambda}
= e^{i\lambda \hat{p}}$. By contrast, in the polymer Hilbert space the
action of $\hat{U}_{\lambda}$ is not weakly continuous in~$\lambda$,
and hence the basic momentum operator does not exist.

The states in the polymer Hilbert space can be described as points in a
certain compact space, the (Harald) Bohr compactification of the real line,
and the operators introduced above can be described in terms of a
representation of the Weyl algebra associated with the classical position
and momentum variables \cite{velhinho,corduneau,hewitt-ross}. There exists
also a `dual' quantization in which a momentum operator and a
family of translation operators in the momenta exist, but there is no basic
position operator~\cite{halvorson}. These mathematical structures will
however not be used in the rest of the paper.

As the basic momentum operator cannot be defined, any phase space function
containing the classical momentum~$p$, most importantly the Hamiltonian,
has to be quantized in an indirect way. With a length scale $\lambda>0$ we
can define the operators
\begin{subequations}
\label{eq:p-and-psquared}
\begin{align}
\hat{p}_{\lambda} &= \frac{1}{2i \lambda} (\hat{U}_\lambda -
\hat{U}^{\dagger}_{\lambda} ) ,
\\
\hat{p}_{\lambda}^2&=
\frac{1}{4\lambda^2} ( 2 - \hat{U}_{2\lambda} -
\hat{U}^{\dagger}_{2\lambda} ) .
\end{align}
\end{subequations}
In $L_2(\mathbb{R})$, the $\lambda\to0$ limit in (\ref{eq:p-and-psquared})
would give the usual momentum and momentum-squared operators $i\partial_x$
and~$-\partial^2_x$. In the polymer Hilbert space the $\lambda\to0$
limit does not exist, and $\lambda$ is regarded as a fundamental length
scale. 
The Hamiltonian operator that corresponds to the classical Hamiltonian
$H = \frac12 p^2 + V(x)$ is then
\be
\hat{H} = \frac{1}{8\lambda^2} ( 2- \hat{U}_{2\lambda} -
\hat{U}^{\dagger}_{2\lambda} ) + \hat{V} ,
\label{H}
\ee
where $V$ is assumed so regular that $\hat{V}$ can be defined by
pointwise multiplication, $\bigl(\hat{V} \psi \bigr) (x) =  V(x)
\psi(x)$. One expects the polymer dynamics to be well approximated by the
Schr\"odinger dynamics in an appropriate regime, and certain results to
this effect are known 
\cite{polymer,fredenhagen-reszewski,corichi-vuka-zapata,Hossain:2009vd}.

Let us now turn to the derivation of the uncertainty relation.  We apply the general result that for two
operators $\hat{A}$ and $\hat{B}$, 
\be
(\Delta A)^2 (\Delta B)^2 \ge \frac{1}{4} \left|\langle [\hat{A},\hat{B}]
\rangle \right|^2 ~,
\label{gen-up}
\ee
where $(\Delta A)^2 \equiv \langle \hat{A} - \langle \hat{A}
\rangle\rangle^2$ and $(\Delta B)^2 \equiv \langle \hat{B} - \langle
\hat{B} \rangle\rangle^2$ are the standard uncertainties in the
measurement of operators in a given state. This identity of course
holds irrespective of the quantization scheme.

Let us denote the polymer basis states as
$|\psi_x\rangle = |x\rangle$. A general normalized state can then 
be expressed as 
\begin{equation}\label{GeneralPolymerState}
    |\Psi\rangle = \sum_{k\,\in\,\mathbb{Z}} c(x_k) |x_k \rangle, 
    \quad \sum_{k\,\in\,\mathbb{Z}} c(x_k)^* c(x_k) = 1 ~.
\end{equation}
It follows from the definition of the momentum operator
(\ref{eq:p-and-psquared}) that
\begin{align}
    \langle [\hat{x},\hat{p}_\lambda ] \rangle & = \frac{1}{2i} \sum_{k\,\in\,\mathbb{Z}}
    c(x_k)^* \left[ c(x_{k}+\lambda) + c(x_{k}-\lambda) \right], \\
    \langle \hat{p}_{\lambda/2}^2 \rangle & =
    \frac{1}{\lambda^2} \sum_{k\,\in\,\mathbb{Z}} c(x_k)^* \left[ 2c(x_k) - c(x_{k}+\lambda)
    - c(x_{k}-\lambda)
    \right].\label{p2}
\end{align}
Combining these two formulae with the normalization condition
(\ref{GeneralPolymerState}) gives the exact expression
\begin{equation}
    \langle [\hat{x},\hat{p}_\lambda ] \rangle = \frac{1}{i} \left[1
    - \frac{\lambda^2}{2} \langle \hat{p}_{\lambda/2}^2 \rangle \right]. 
\ee    
It is interesting to note that the {\it rhs} vanishes for states that satisfy $ \langle \hat{p}_{\lambda/2}^2\rangle = 2/\lambda^2$. Expanding for small $\lambda$ gives 
\begin{equation}\label{up-small-lambda-gen}
      \Delta x \Delta p_\lambda \ge \frac{1}{2} \left[ 1
    - \frac{\lambda^2}{2} \langle \hat{p}_{\lambda}^2 \rangle +
    \mathcal{O}(\lambda^4) \right].
\end{equation}
These results hold for any normalizable state. For  states such that $\langle
\hat{p}_\lambda\rangle=0$, the last equation becomes
\be
\quad \Delta x \Delta p_\lambda \ge \frac{1}{2} \left[ 1
    - \frac{\lambda^2}{2} (\Delta{p}_{\lambda})^2   
    + \mathcal{O}(\lambda^4) \right],
\ee 
which is exactly of the form (\ref{mup}) that arises from
the gedanken arguments and  string theory. It is worthwhile stressing
that this result does not rely on any sort of uniform sampling
approximation where one assumes $x_{k+1} - x_k = \lambda$.

We note that the sign of the correction term is \emph{negative} definite, therefore  the
position-momentum uncertainty   decreases due to the presence of
the scale $\lambda$.
Secondly, the polymer case is evidently a {\it kinematic} computation, whereas
the string argument requires  the dynamical input that  the size of a string at
high energy depends  on its energy.

In Schr\"odinger quantization minimum uncertainty states are the Gaussian
states, therefore  it is instructive to see the polymer corrections for the 
Gaussian state 
\be
\label{QMGaussianState}
|\Psi(x_0,p_0;\sigma)\rangle = \frac{1}{\mathcal{N}}
\sum_{k\,\in\,\mathbb{Z}} e^{-(x_k-x_0)^2/2\sigma^2}
e^{-ip_0x_k} |x_k\rangle ~,
\ee
where $\mathcal{N}$ is the normalization factor. Using  actions of the
operators $\hat{x}$, $\hat{p}_{\lambda}$ on this  state,  computation of the \emph{rhs} of  
(\ref{gen-up}) gives
 \be
 \Delta x^2\Delta {p_\lambda}^2 \ge \frac{1}{4} 
  e^{-\lambda^2/2\sigma^2}  \cos^2\lambda p_0 ~,
 \label{Pup}
 \ee
where we have chosen uniform sampling and approximated sums by integrals. (The expectation values can be computed without   this approximation by using the Poisson resummation
formula, as used for example in \cite{HW-cosm}.) In the regime where
$\lambda p_0 \ll 1$, 
$\lambda^2/4\sigma^2 \ll 1$, and using the fact that $\langle
\hat{p}_\lambda\rangle \approx p_0$ and $\Delta {p_{\lambda}^2} \approx
1/2\sigma^2$, the equation (\ref{Pup}) gives
\be
 \Delta x\Delta {p_\lambda} \ge \frac{1}{2} \left[1 - \frac{\lambda^2}{2}
\left(\langle \hat{p}_\lambda\rangle^2  + (\Delta{p_\lambda})^2 \right) +
\mathcal{O}(\lambda^4) \right] ~,
\ee
which is same as  eqn. (\ref{up-small-lambda-gen}).

\section{Uncertainty relation in polymer field theory}

We consider in this section a generalization to field theory  \cite{HHS-cosm} of the quantization  described above. Our aim is to demonstrate how the
uncertainty relation acquires a metric dependence. Let us focus on  scalar
field with canonical variables $(\phi(x,t), P(x,t))$ on a background
\be
ds^2  = -dt^2 + q_{ab}dx^a dx^b,
\ee
where $q_{ab}$ is the spatial metric.  The classical polymer variables we
consider are
\begin{equation}
\label{BasicVariables}
\phi_f \equiv \int d^3x \sqrt{q} \, f(x) \phi(x) ,
\quad U_{\lambda} \equiv \exp\left( \frac{i \lambda
P }{\sqrt{q}} \right),
\end{equation}
where the smearing function $f(x)$ is a scalar. Since $P $ transforms as a
density under coordinate transformations, the $\sqrt{q}$ factor in the
second definition is required to make the argument of the exponent a
scalar. The parameter $\lambda$ is now a spacetime constant with dimensions
of $(\text{mass})^{-2}$.  These  variables satisfy the Poisson algebra
\begin{equation}
\label{basic-pb} \{\phi_f,U_\lambda\} = if\lambda U_\lambda.
\end{equation}

Now having seen how the metric enters  in the definition of these variables,
let us  specialize to an Friedman-Robertson-Walker  spacetime
with spatial metric $q_{ab} = a^2(t) q^0_{ab}$, where $a(t)$ is the scale
factor and the fiducial metric $q^0_{ab} = \delta_{ab}$. To keep it
compatible with homogeneity, we set the smearing function $f(x) = 1$.
Employing a standard box normalization to regulate the spatial integration
in (\ref{BasicVariables}) gives  the  reduced variables 
\begin{equation}\label{ReducedVariables}
\phi_f = V_0 a^3 \phi , \quad U_\lambda= \exp\left(i\lambda
P/a^3 \right),
\end{equation}
where $V_0 = \int d^3 x \sqrt{q^0}$ is the fiducial volume. The
Poisson brackets of the reduced variables is the same as that in eqn. 
(\ref{basic-pb}) with $f=1$. Quantization proceeds as before by realizing
this Poisson algebra  as a commutator algebra on the polymer Hilbert space:
\begin{equation}\label{operator actions}
  \hat\phi_f |\mu\rangle = \mu|\mu\rangle,
  \quad \hat U_{\lambda} |\mu \rangle = |\mu + \lambda\rangle ~.
\end{equation}
The scale dependent momentum is now
\begin{equation}\label{SFMomentum}
P_\lambda= \frac{a^3}{2i\lambda}
(U_{\lambda} - U^{\dagger}_{\lambda }) ~,
\end{equation}
where we note the non-trivial scale factor dependence. To derive the
uncertainty relations we again choose a Gaussian state peaked at the phase
space values $(\phi_0,P_0)$:
\begin{equation}
\label{GaussainStateFRW}
    |\Psi\rangle = \frac{1}{\mathcal{N}} \sum_{k\,\in\,\mathbb{Z}} c_k
    |\lambda_k\rangle, \quad
    c_k \equiv e^{-(\phi_k-\phi_0)^2/2\sigma^2} e^{-iP_0 \phi_k V_0},
\end{equation}
where $\phi_k = \lambda_k/V_0 a^3$ is an eigenvalue of the scalar field
operator derived from $\hat{\phi}_f$ in eqn.~(\ref{operator actions}).
The scalar configuration points $\lambda_k$ are chosen
such that the Gaussian profile is well sampled. Here we choose a uniform
sampling. This state gives the expectation value
\begin{equation}\label{U EV}
    \langle \hat{U}_{\lambda} \rangle = e^{i\Theta}
    e^{-\Theta^2/4\Sigma^2},
\end{equation}
where
\begin{equation}
    \Theta \equiv \lambda P_0 a^{-3}, \ \ \ \
\quad \Sigma \equiv \sigma
    V_0 P_0.
\end{equation}
are dimensionless variables.
By computing the \emph{rhs} of the equation (\ref{gen-up}) for the 
Gaussian state (\ref{GaussainStateFRW}), we 
arrive at the scale factor dependent uncertainty relation
\begin{equation}
\label{FRWUncertaintyRelationRHS}
V_0^2 \left(\Delta \phi\right)^2 \left(\Delta P_\lambda\right)^2
 \ge \frac{1}{4} e^{-\Theta^2/2 \Sigma^2} \cos^2\Theta ~.
\end{equation}
This  is very similar to the equation (\ref{Pup}); the $V_0$ factor on the  {\it rhs} is present due to
the reduced classical Poisson bracket $\{\phi, P\}=1/V_0$.

Computing the \emph{lhs} of  eqn. (\ref{gen-up}) explicitly for the Gaussian state, we find that
expectation values of the smeared field operators are
\be
\label{SmearedFieldEV}
\langle \hat{\phi_f} \rangle =  \phi_0 V_0 a^3, \ \ \ \ \ \ \
\langle \hat{\phi_f^2} \rangle =  \phi_0^2 V_0^2a^6 +
\frac{1}{2}V_0^2a^6\sigma^2,
\ee
which give the field fluctuation
\be
(\Delta \phi)^2 \equiv \frac{1}{V_0^2a^6} \left(
\langle\hat{\phi_f^2}\rangle - \langle\hat{\phi_f}\rangle^2\right)
= \frac{\sigma^2}{2} ~.
\ee
Similarly, the expectation value of the field momentum operators are
\begin{equation}
\label{FieldMomentumEV}
\langle \hat{P}_\lambda \rangle = \frac{a^3}{\lambda}
e^{-\Theta^2/4\Sigma^2} \sin\Theta, \ \ \ \ \ \ \ \ \
\langle \hat{P}_{\lambda}^ 2 \rangle = \frac{a^6}{2\lambda^2}
\left(1 - e^{-\Theta^2/\Sigma^2} \cos 2\Theta\right),
\end{equation}
\begin{equation}
\label{FieldMomentumSqEVDef}
\Delta P^2_\lambda \equiv
\langle \hat{P}_\lambda^2\rangle - \langle\hat{P}_\lambda
\rangle^2 = \frac{a^6}{2\lambda^2}
\left(1 - e^{-\Theta^2/2 \Sigma^2}\right)
\left(1 + e^{-\Theta^2/2 \Sigma^2} \cos 2\Theta\right)
\end{equation}
 Thus the {\it lhs} of the uncertainty relation is 
\begin{equation}
\label{FRWUncertaintyRelation}
V_0^2 \left(\Delta \phi\right)^2 \left(\Delta P_\lambda\right)^2
 = \frac{\Sigma^2}{4\Theta^2}
\left(1 - e^{-\Theta^2/2 \Sigma^2}\right)
\left(1 + e^{-\Theta^2/2 \Sigma^2} \cos 2\Theta\right) ~.
\end{equation}
The scale factor dependence in this equation appears  through the
variable $\Theta$. For large volumes ($a \rightarrow \infty$)  or
vanishing discreteness scale  ($\lambda \rightarrow 0$),  we have
$\Theta \rightarrow 0$.  Therefore it is useful to expand the right hand side
of the equation (\ref{FRWUncertaintyRelation}) for small $\Theta$. This
leads to
\begin{equation}
\label{FRWUPSmallTheta}
V_0 \left(\Delta \phi\right) \left(\Delta P_\lambda \right) =
 \frac{1}{2} - \frac{\Theta^2}{4} \left(1 + \frac{1}{2\Sigma^2}\right)
  = \frac{1}{2} \left[ 1 - \lambda^2 \left[ \frac{1}{2a^6}\left(
  \langle \hat{P}_\lambda \rangle^2 + (\Delta P_\lambda)^2 \right) \right]
  + \mathcal{O}(\lambda^4) \right],
\end{equation}
where we have used the expansion 
 \begin{equation}
    \langle \hat{P}_\lambda^2\rangle  = \frac{a^6}{\lambda^2}\Theta^2\left( 1+\frac{1}{2\Sigma^2} \right)
    + \mathcal{O}\left( \Theta^4\right),
\end{equation}
of (\ref{FieldMomentumEV})  for small $\Theta$. With the  identifications
$a\rightarrow 1$, $P_\lambda \rightarrow p_\lambda$ and $\phi_f \rightarrow
x$, we note that  eqn. (\ref{FRWUPSmallTheta}) becomes  the same as the {\it rhs} of
eqn. (\ref{up-small-lambda-gen}).  This means that with polymer corrections up to order $\lambda^2$, Gaussian states remain minimum uncertainty states. However, this is no longer the case if  the next to leading order corrections are included,  as is apparent from a comparison of the 
\emph{rhs} of eqn. (\ref{FRWUncertaintyRelationRHS}) and eqn. (\ref{FRWUncertaintyRelation}).

For massless free scalar field, the $\lambda^2$ correction in the equation
(\ref{FRWUPSmallTheta}) has a remarkable form: the first term inside
the square brackets is the classical scalar field energy density, and the
second is its fluctuation. Thus we see that the correction to the
uncertainty relation for a massless scalar field propagating in the FRW
background acquires corrections proportional to the energy density.

It also follows from eqn. (\ref{FRWUPSmallTheta}) that in the early
universe, the corrections to the uncertainty principle we have found were much larger. 
The $\Theta \ll 1$ expansion of eqns. (\ref{FRWUncertaintyRelationRHS},
\ref{FRWUncertaintyRelation}) of course breaks downs for   $\Theta\sim 1$,
and  is not  reliable also for sufficiently large values of $\Theta$, where the approximation of 
quantum field on a fixed background breaks down. Nevertheless, it is evident that the modifications
to the uncertainty principle are significant in a certain epoch. This might lead to
some interesting consequences for early universe physics.

\section{Summary}

We have seen that polymer quantization gives a modified  uncertainty
relation that resembles the one coming from  heuristic arguments involving
gravity, from string theory, and from generalized commutators of
position and momentum. The common feature in all these approaches is  
a  length scale in addition to $\hbar$. However it is only
in the polymer approach that the quantization method itself comes with a
scale due to the choice of Hilbert space, and in this sense the
modifications are independent of the theory.

More generally we have seen that if polymer quantization is applied to
field theory on a given background, the uncertainty relation acquires an
explicit metric  dependence. This dependence  is dramatically illustrated
for the case of the Friedmann-Robertson-Walker  background, where we see
that the correction dominates at early times. This feature could lead to
interesting consequences in the physics of early universe. It would also be
of interest to see what happens for other backgrounds, especially the black
hole, to see how the presence of an event horizon affects the relation. We
note finally that Yoneya's  general arguments \cite{Yoneya} that
$\Delta p$ corrections to the $\Delta x \Delta p$ uncertainty relation
give rise to a position-time uncertainty relation $\Delta x \Delta t \ge l_P^2$ apply 
equally well in the setting we have discussed.  Therefore polymer
quantization gives an indication of spacetime non-commutativity. This is
another intriguing aspect that deserves further study.
 
\vskip 0.3cm.

\noindent{\bf Acknowledgments} This work was supported in part by NSERC of
Canada, and by the Atlantic Association for Research in the mathematical
Sciences (AARMS).

\bibliographystyle{apsrev}
\bibliography{polymer-UP}

\end{document}